\documentclass[a4paper,11pt]{article}

\usepackage{contribution}

\newcommand{\weblink}[2][]{%
    \ifthenelse{\equal{#1}{}}%
    {\textnormal{\url{#2}}}%
    {\textnormal{\href{#2}{#1}}}%
}

\def\beq{\begin{equation}}
\def\eeq#1{\label{#1}\end{equation}}
\def\eeqn{\end{equation}}

\def\beqa{\begin{eqnarray}}
\def\eeqa#1{\label{#1}\end{eqnarray}}
\def\eeqan{\end{eqnarray}}

\let\bar=\overbar

\def\etal{{\it et al.}}

\def\Dslash{\not{\hbox{\kern-4pt $D$}}}
\def\dslash{\not{\hbox{\kern-2pt $\del$}}}

\def\msb{{\bar{\ssstyle M \kern -1pt S}}}

\newcommand{\contribution}[7][]{%
  \clearpage
  \thispagestyle{plain}
  \ifthenelse{\equal{#1}{}}
  {\hypersetup{pdftitle={#2}}}
  {\hypersetup{pdftitle={#1}}}
  \hypersetup{pdfauthor={{#3} {#4}}}
  {\centering\normalfont\LARGE\bfseries\sffamily #2 \par\nobreak}
  \lhead{}
  \chead{%
    \textit{\footnotesize XIV International Conference on Hadron Spectroscopy
      (\weblink[\textit{hadron2011}]{http://www.hadron2011.de}), 13-17 June 2011, Munich, Germany}%
  }
  \rhead{}
  \bigskip
  \begin{center}
    {#3} {#4}\ifthenelse{\equal{#6}{}}{}{\footnote{\weblink[#6]{mailto:#6}}}
    \ifthenelse{\equal{#7}{}}{}{#7} \\
    \textit{#5}
  \end{center}
  \bigskip
}

\renewcommand{\abstract}[1]{%
  \begin{center}
    \begin{minipage}{0.85\textwidth}
      \begin{footnotesize}
        #1
      \end{footnotesize}
    \end{minipage}
  \end{center}
  \bigskip
}

\begin{document}

%
%
%
%
%
{  


%

\contribution[Quarkonium production in pp collisions at 7 TeV with the CMS experiment]  
{Quarkonium production in pp collisions at 7 TeV with the CMS experiment}  
{Bora}{Akg\"{u}n}  
{Department of Physics \\
  Carnegie Mellon University \\ 
  5000 Forbes Avenue, Pittsburgh, PA 15213, USA }  
{bora.akgun@cern.ch}  
{on behalf of the CMS Collaboration}  
%

\abstract{%
  The production of $J/\psi$ and $\Upsilon$ mesons is studied in pp collisions at $\sqrt s = 7$ TeV with the CMS experiment at the LHC. 
  The $J/\psi$ measurement is based on a dimuon sample corresponding to an integrated luminosity of 314~nb$^{-1}$. The $J/\psi$ differential cross section is determined, as a function of the $J/\psi$ transverse momentum, in three rapidity ranges.  A fit to the decay length distribution is used to separate the prompt from the non-prompt (b hadron to $J/\psi$) component. Integrated over the $J/\psi$ transverse momentum from 6.5 to 30~GeV/$c$
and over rapidity in the range $|y| < 2.4$, the measured cross sections, times the dimuon decay branching fraction, are $70.9 \pm 2.1 (\mbox{stat.}) \pm 3.0 (\mbox{syst.}) \pm 7.8 (\mbox{lumi.})$~nb for prompt $J/\psi$ mesons, assuming unpolarized production, and $26.0 \pm 1.4 (\mbox{stat.}) \pm 1.6 (\mbox{syst.}) \pm 2.9 (\mbox{lumi.})$~nb for $J/\psi$ mesons from b-hadron decays. 
The $\Upsilon$ measurement is based on a dimuon sample corresponding to an integrated luminosity $3.1\pm 0.3\,\text{pb}^{-1}$. Integrated over the rapidity range $|y|<2$, we find the product of the $\Upsilon(1S)$ production cross section and branching fraction to dimuons to be $7.37 \pm 0.13(\mbox{stat.})^{+0.61}_{-0.42} (\mbox{syst.}) \pm 0.81 (\mbox{lumi.})$ \,nb.  This cross section is obtained assuming unpolarized $\Upsilon(1S)$ production. If the $\Upsilon(1S)$ production polarization is fully transverse or fully longitudinal, the cross section changes by about 20\,\%. 
}

The hadroproduction of quarkonia is not understood since none of the existing theories successfully reproduces both the differential cross section and the polarization measurements of the $J/\psi$ or $\Upsilon$ states~\cite{yellow}. It is expected that studying quarkonium hadroproduction at higher center-of-mass energies and over a wide rapidity range will facilitate significant improvements in our understanding. Measurements of quarkonium hadroproduction cross sections and production polarizations made at the Large Hadron Collider (LHC) will allow important tests of several alternative theoretical approaches, including non-relativistic QCD (NRQCD) factorization~\cite{bib-nrqcd}.

The data samples used in these analyses were recorded by the CMS detector~\cite{JINST} in $pp$ collisions at a center-of-mass energy of 7 TeV. Many more details in these analyses are available in Refs.~\cite{paper1,paper2}. The trigger requires the detection of two muons at the hardware level, without any further selection at the high-level trigger (HLT). The coincidence of two muon signals without an explicit p$_{T}$ requirement is sufficient to maintain the dimuon trigger without prescaling. All three detectors in the muon systems, drift tubes, cathode-strip and resistive-plate chambers, take part in the trigger decision.

Each muon is required to satisfy:
\begin{align}
& p_{T}^{\mu}> 3.5~\mbox{GeV/c}~~\mbox{if}~~|\eta^{\mu}| < 1.6 ; \\
\nonumber
& p_{T}^{\mu} > 2.5~\mbox{GeV/c}~~\mbox{if}~~1.6 < |\eta^{\mu}| < 2.4~~~\mbox{(for}~\Upsilon~\mbox{selection)} . \\
& p_{T}^{\mu}> 3.3~\mbox{GeV/c}~~\mbox{if}~~|\eta^{\mu}| < 1.3;~~p^{\mu} > 2.9~\mbox{GeV/c}~~\mbox{if}~~1.3 < |\eta^{\mu}| < 2.2 ;\\ 
\nonumber
& p_{T}^{\mu} > 2.4~\mbox{GeV/c}~~\mbox{if}~~2.2 < |\eta^{\mu}| < 2.4~~~\mbox{(for}~J/\psi~\mbox{selection)} .
\end{align}
The dimuon invariant-mass spectra in the J/$\psi$ and $\Upsilon(nS)$ regions are shown in Fig.~\ref{fig:mass}. 
\begin{figure}[htb]
  \begin{center} $
  \begin{array}{cc}
    \includegraphics[height=4.cm]{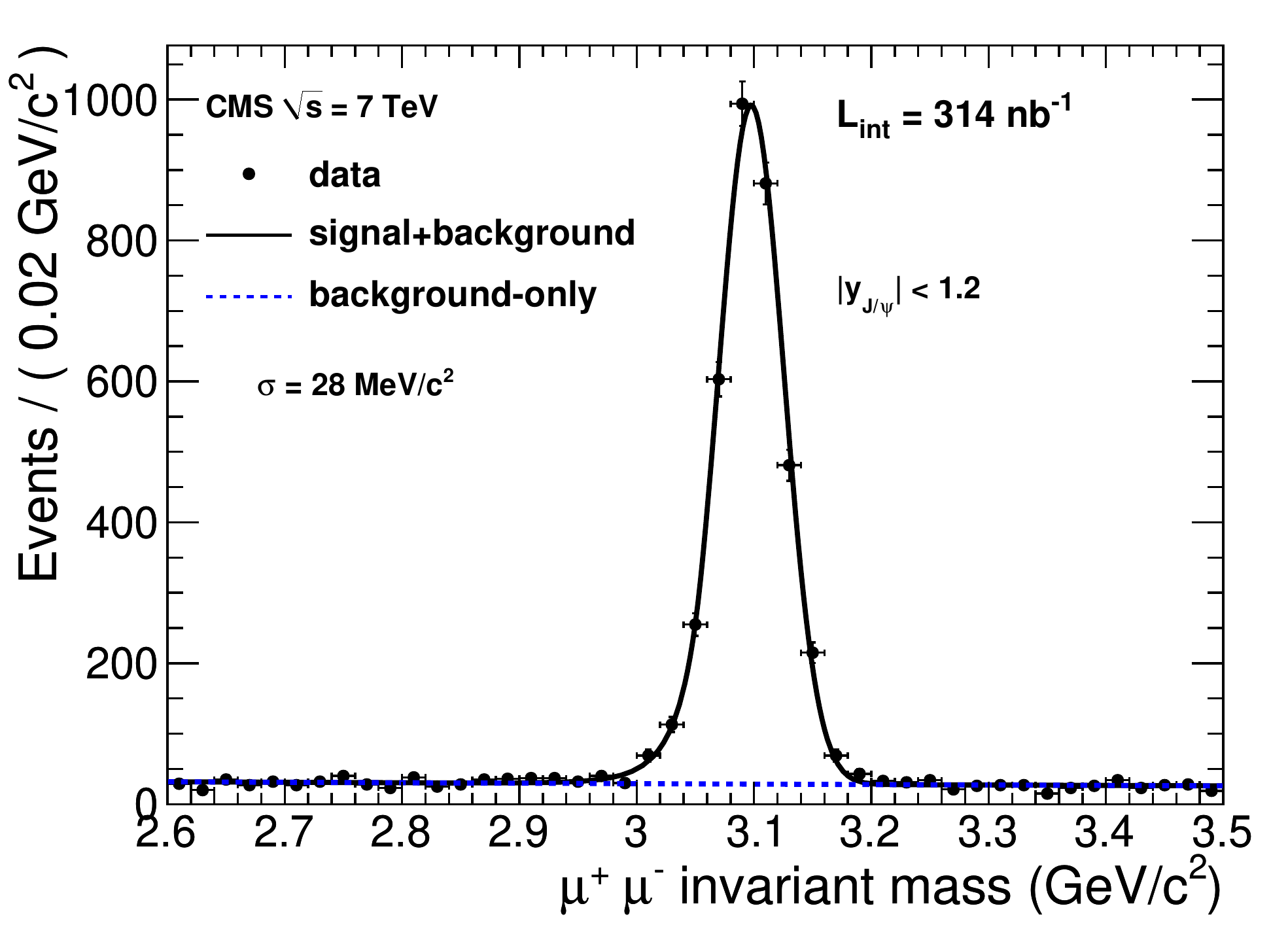} &
    \includegraphics[height=4.cm]{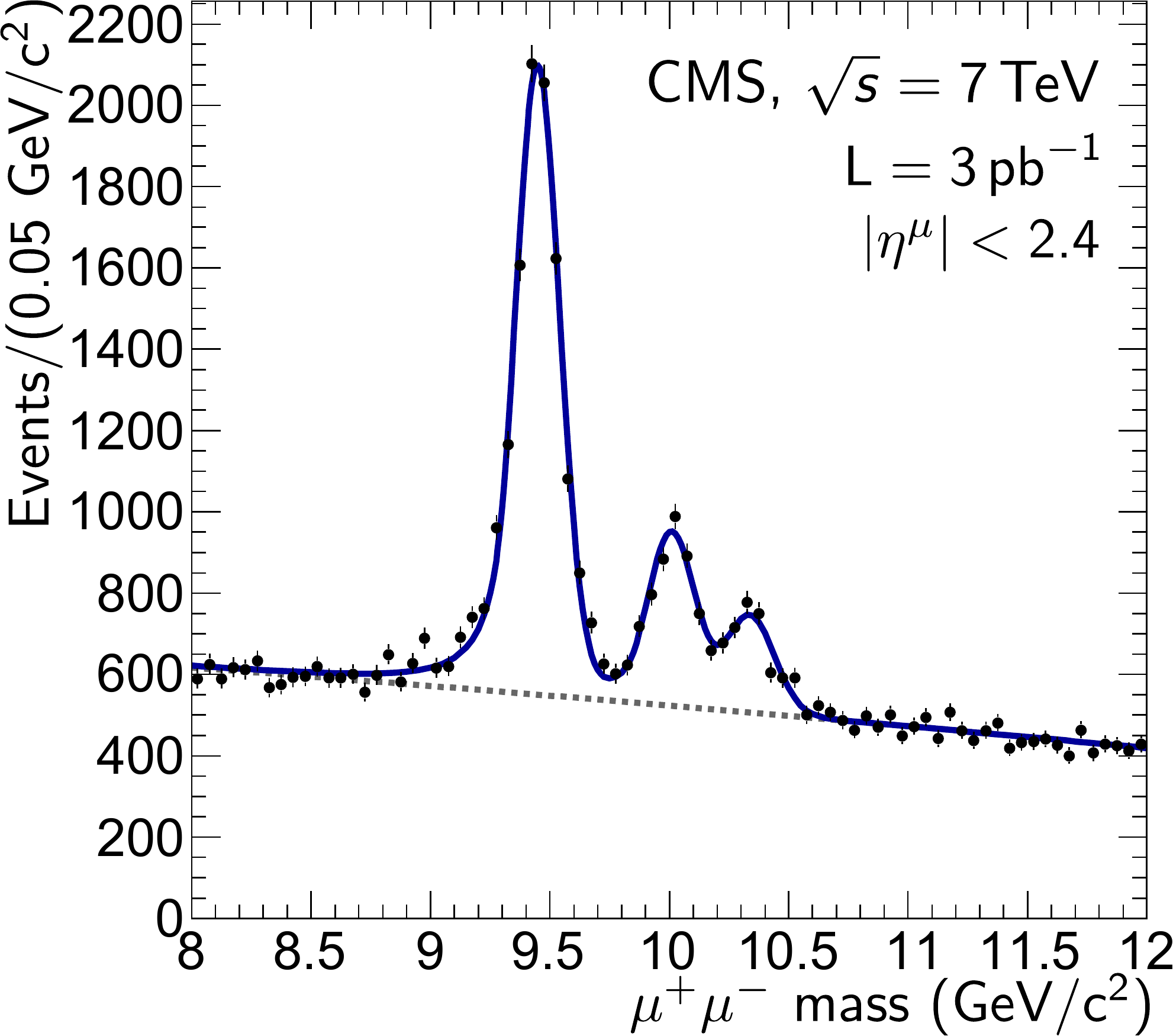}
   \end{array}$
    \caption{The dimuon invariant-mass distribution in the vicinity of the J/$\psi$ resonance for $|y_{J/\psi}| < $1.2 (left) and $\Upsilon(nS)$ resonances for $|\eta^{\mu}| < $2.4 (right) .}
    \label{fig:mass}
  \end{center}
\end{figure}
The acceptance $A$ is defined as the fraction of detectable $J/\psi \to\mu^+\mu^-$ or $\Upsilon\to\mu^+\mu^-$ decays, 
as a function of the dimuon transverse momentum p$_{T}$ and rapidity $y$,
\begin{equation}
A(p_{T}, y;\lambda_\theta)=\frac
{N_{\mbox{\small{det}}}(p_{T}, y;\lambda_\theta)}
{N_{\mbox{\small{gen}}}(p_{T}, y;\lambda_\theta)} \quad ,
\end{equation}
where $N_{\mbox{\small{det}}}$ is the number of detectable J/$\psi$ or $\Upsilon$~events in a given (p$_{T}$, $y$) bin, expressed in terms of the dimuon variables after detector smearing, and $N_{\mbox{\small{gen}}}$ is the corresponding total number of generated J/$\psi$ or $\Upsilon$~events in the Monte Carlo simulation. The parameter $\lambda_\theta$ reflects the fact that the acceptance is computed for various polarization scenarios.

The total muon efficiency can be factorized into three conditional terms,
\begin{equation}
  \epsilon ({\rm total}) = \epsilon ({\rm trig | \rm id}) \cdot \epsilon({\rm id
    | track}) \cdot \epsilon({\rm track | accepted}) \equiv \epsilon_{\rm
    trig} \cdot \epsilon_{\rm id} \cdot \epsilon_{\rm track}\,.
\end{equation}
The tracking efficiency, $\epsilon_{\rm track}$, is the efficiency for a muon track from the quarkonium decay to be reconstructed in the presence of other activity in the silicon tracker, as determined with a track-embedding technique~\cite{bib-trackingefficiency}. The muon identification efficiency, $\epsilon_{\rm id}$, is the probability that the track in the silicon tracker is identified as a muon and is based on the tag-and-probe~\cite{bib-muonreco} method. The efficiency that an identified muon satisfies the trigger, $\epsilon_{\rm trig}$, is again measured with the same technique.

The differential cross section is determined from the signal yield, $N_{\rm{fit}}$,  obtained directly from a weighted fit to the dimuon invariant-mass spectrum, after correcting for the acceptance (${\cal{A}}$) and the total efficiency ($\epsilon$), through the equation:
\begin{equation}
      \frac{ d^{2}\sigma } { dp_{T} \, dy} \cdot {\cal B} \left(X\rightarrow\mu^+\mu^-\right)  =
      \frac{ N_{\text{fit}} \langle \frac{1}{\mathcal A, \epsilon} \rangle}
      {{\mathcal L} \ \cdot \Delta p_{T} \cdot \Delta y }\,,
\end{equation}
upon normalization by the integrated luminosity of the dataset, $\mathcal{L}$, and by the bin widths, $\Delta p_{T}$ and $\Delta y$, of the dimuon transverse momentum and rapidity.

The yields are extracted via an unbinned maximum likelihood fit. The measured mass-lineshape of the J/$\psi$ and each of the three $\Upsilon$ states is parameterized by a ``Crystal Ball'' (CB) function. A second-degree polynomial is chosen to describe the background for the $\Upsilon(nS)$ states and an exponential function for the J/$\psi$ background.  

The $\Upsilon(nS)$ integrated production cross sections times branching fractions for the range $|y|<2$ are measured to be:
\begin{align}
   \sigma(pp \rightarrow \Upsilon(1S) X ) \cdot {\cal B} (\Upsilon(1S)  \rightarrow \mu^+
    \mu^-) &= 7.37 \pm 0.13 (\rm stat.)^{+0.61}_{-0.42} (\rm syst.)\pm 0.81 (\rm lumi.)\,\text{nb}\,, \\
    \nonumber
    \sigma(pp \rightarrow \Upsilon(2S) X ) \cdot {\cal B} (\Upsilon(2S) \rightarrow \mu^+
    \mu^-) &= 1.90 \pm 0.09 (\rm stat.)^{+0.20}_{-0.14} (\rm syst.)\pm 0.24 (\rm lumi.)\,\text{nb}\,, \\
    \nonumber
    \sigma(pp \rightarrow \Upsilon(3S) X ) \cdot {\cal B} (\Upsilon(3S) \rightarrow \mu^+
    \mu^-) &= 1.02 \pm 0.07 (\rm stat.)^{+0.11}_{-0.08} (\rm syst.) \pm 0.11 (\rm lumi.)\,\text{nb}\,.
  \end{align}
The $\Upsilon(1S)$ and $\Upsilon(2S)$ measurements include feed-down from higher-mass states, such as the $\chi_b$ family and the $\Upsilon(3S)$. These measurements assume unpolarized $\Upsilon(nS)$ production. Assumptions of fully-transverse or fully-longitudinal polarizations change the cross sections by about 20\%. The $\Upsilon(nS)$ differential p$_{T}$ cross sections are shown in Fig.~\ref{fig:xsec_ptdiff} (left).

\begin{figure}[htb]
  \begin{center} $
  \begin{array}{ccc}
    \includegraphics[height=4.02cm]{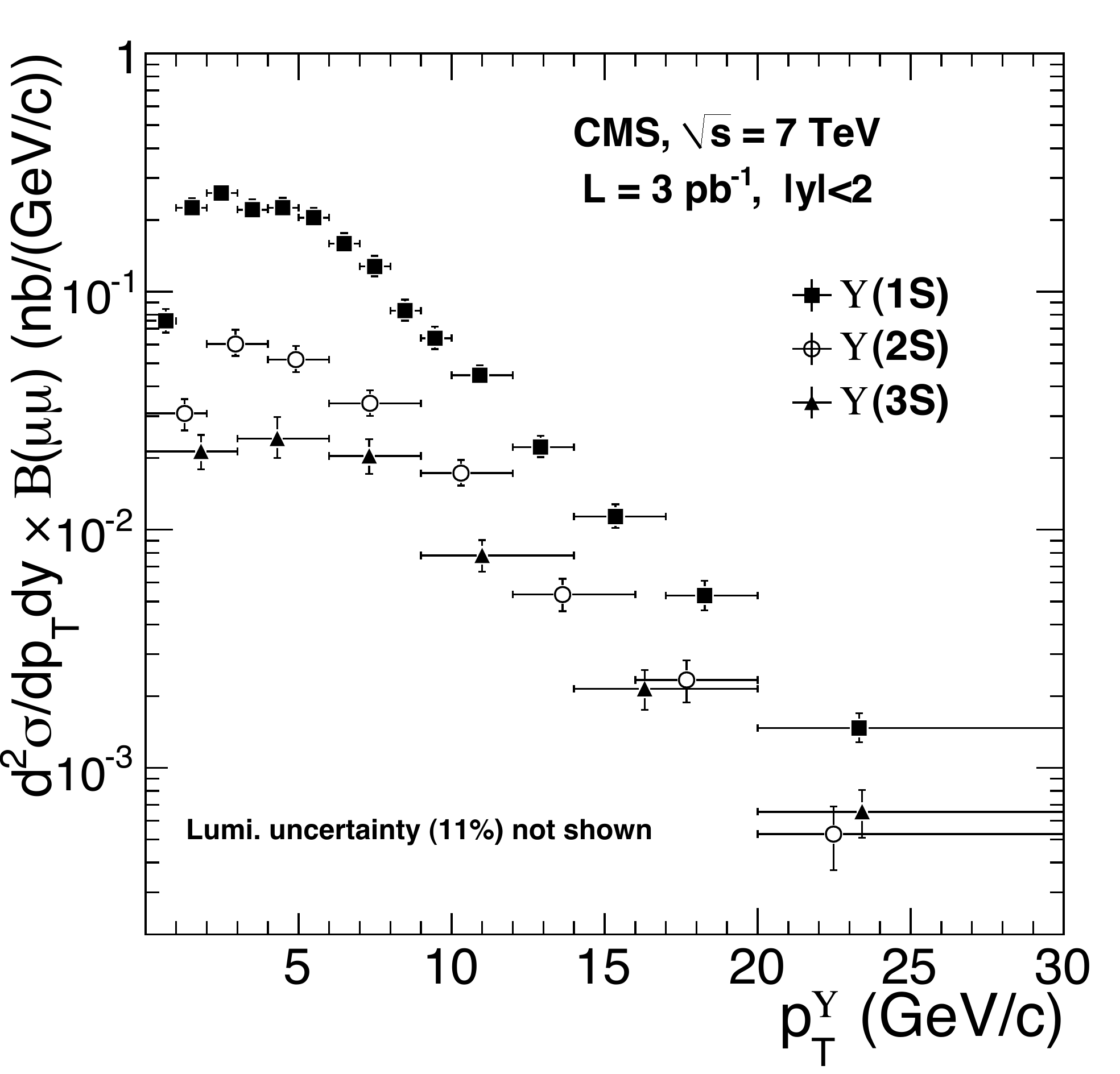} &
    \includegraphics[height=4.02cm]{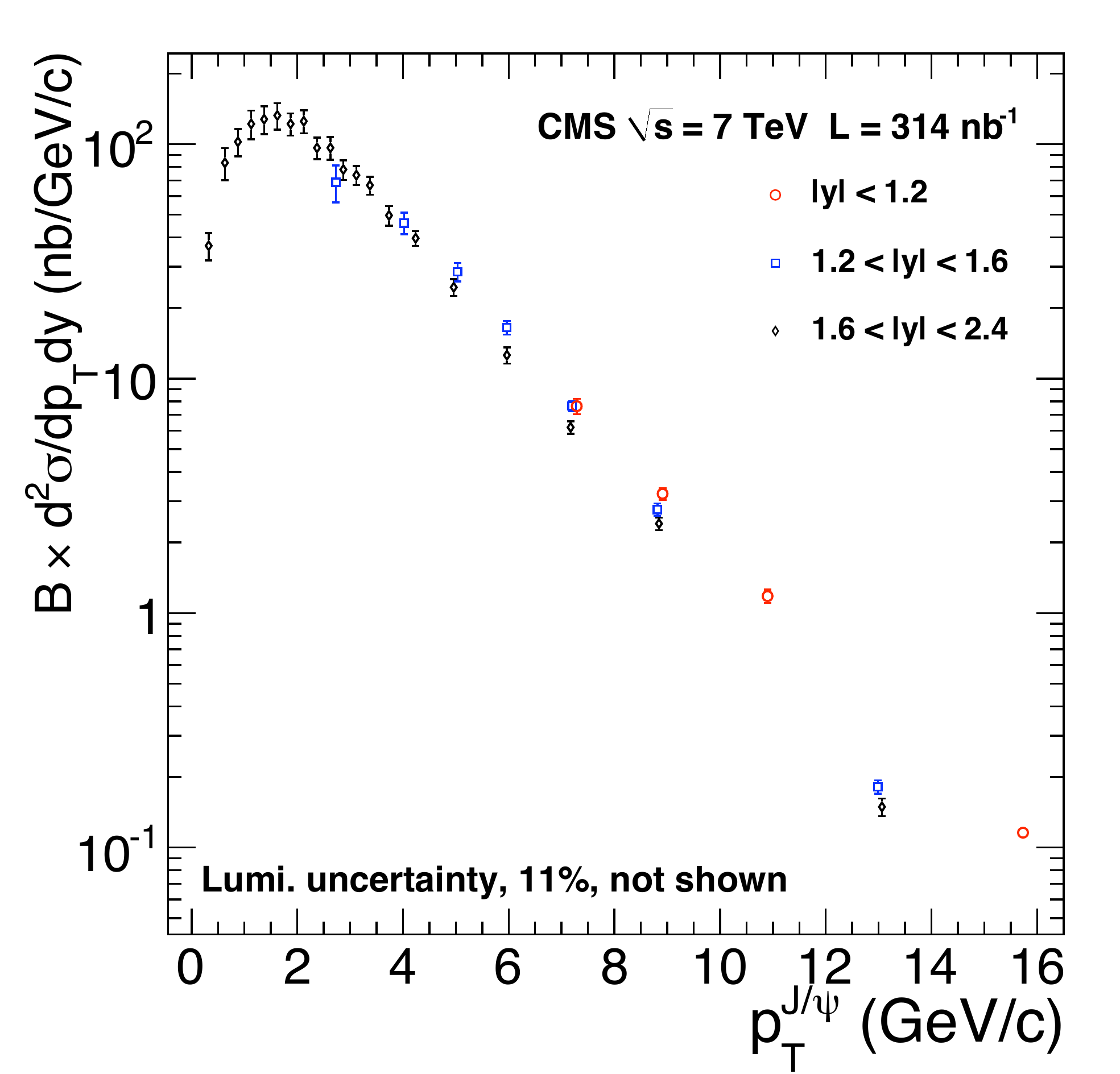} &
    \includegraphics[height=4.02cm]{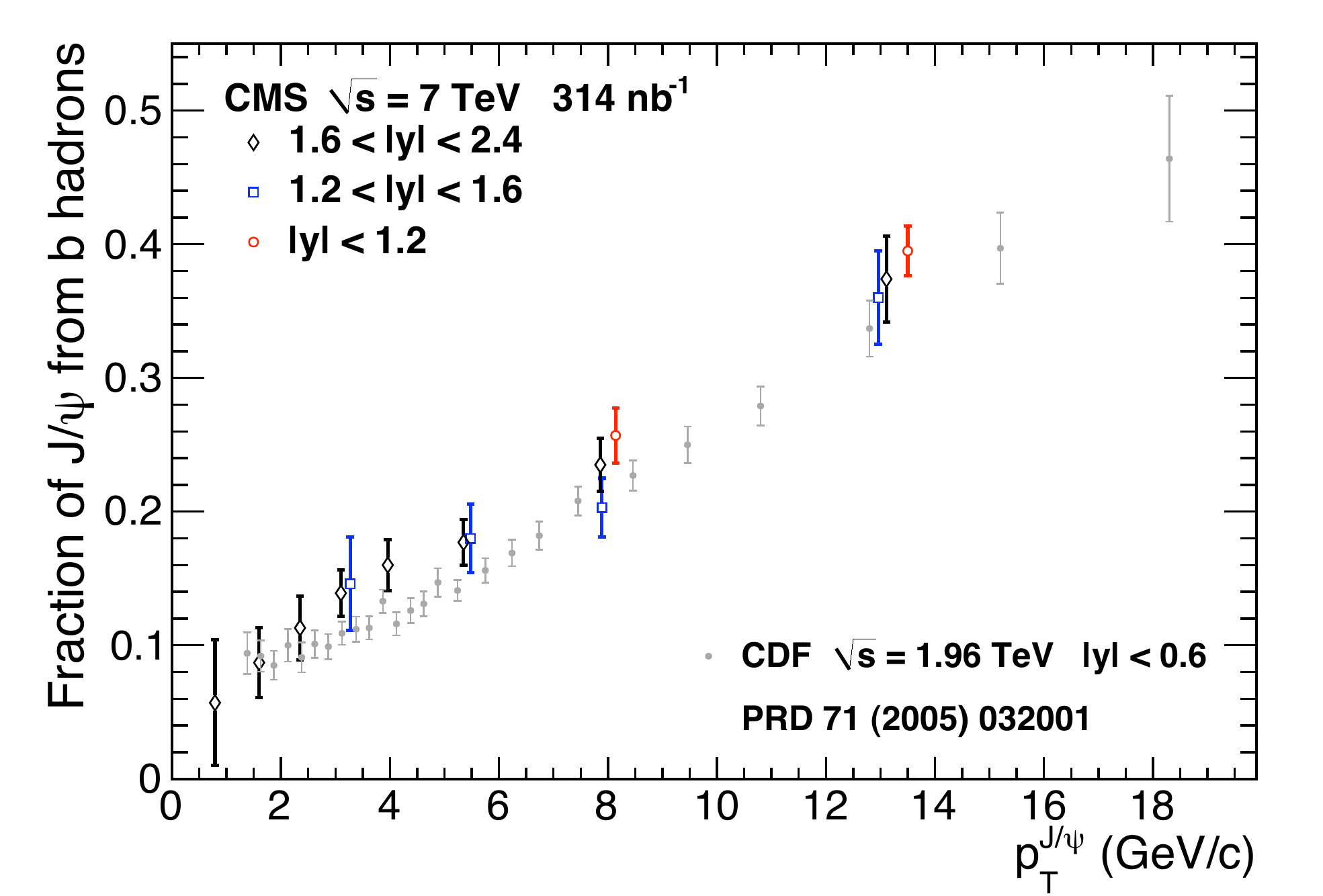}
   \end{array}$
    \caption{(left) $\Upsilon(nS)$ differential cross sections in the rapidity interval $|y|<2$, (middle) fraction of the J/$\psi$ production cross section originating from b-hadron decays, as a function of the J/$\psi$ p$_{T}$, as measured by CMS in three rapidity bins and by CDF, at a lower collision energy, (right) differential inclusive $J/\psi$ cross section as a function of $p_{T}$ for the three different rapidity intervals and in the unpolarized-production scenario.}
    \label{fig:xsec_ptdiff}
  \end{center}
\end{figure}

The total cross section times branching fraction for inclusive J/$\psi$~production, obtained by integrating over p$_{T}$ between 6.5 and 30 GeV/c and over rapidity $|y| <$ 2.4 in the unpolarized-production hypothesis, is:
 \begin{equation}
{\small \sigma(pp \rightarrow J/\psi +X)\cdot\mbox{BR}(J/\psi\to\mu^+\mu^-) = 97.5 \pm 1.5\mbox{(stat.)} \pm 3.4\mbox{(syst.)} \pm 10.7\mbox{(lumi.)} \mbox{ nb} .}
 \end{equation}
Fig.~\ref{fig:xsec_ptdiff} (middle) shows the inclusive differential cross section in the three rapidity ranges, and Fig.~\ref{fig:xsec_ptdiff} (right) shows the measured fraction of J/$\psi$ from b-hadron decays (non-prompt) as a function of J/$\psi$ p$_{T}$. It increases strongly with p$_{T}$. At low p$_{T}$, essentially all J/$\psi$ mesons are promptly produced, whereas at p$_{T}$ $\sim$\,12 GeV/c around one third come from beauty decays. This pattern does not show a significant change with rapidity (within the current uncertainties) over the range covered by the CMS detector. In Fig.~\ref{fig:xsec_ptdiff} (right), the CMS results are compared to the measurements of CDF~\cite{bib-cdfjpsi}, obtained in $p\bar{p}$ collisions at $\sqrt{s} = 1.96$~TeV. It is interesting to note that the increase of the b fraction with p$_{T}$ is very similar between the two experiments, the CMS points being only slightly higher, despite the different collision energies.

The J/$\psi$ differential prompt and non-prompt measurements have been compared
with theoretical calculations. A reasonable agreement is found between data and theory
for the non-prompt case while the measured prompt J/$\psi$ cross section exceeds the
expectations at forward rapidity and low $p_{T}$.

Under the assumption that the cross section is uniform in rapidity for the measurement range of each experiment, the $\Upsilon(nS)$ cross sections measured at $\sqrt{s}$ =7 TeV are about three times larger than the cross sections measured at the Tevatron. This work provides new experimental results which will serve as input to ongoing theoretical investigations of the correct
description of bottomonium production. 


%

}  


\end{document}